\documentclass[a4paper,pra,twocolumn,showpacs,superscriptaddress]{revtex4-1}

\usepackage{epsfig}
\usepackage{amsbsy,latexsym}
\usepackage{amsmath}
\usepackage{amsthm}
\usepackage{amssymb, mathrsfs}
\usepackage[mathscr]{eucal}
\usepackage{color}
\usepackage{subfigure}
\usepackage[normalem]{ulem}
\usepackage{cancel}

\def\Tr{\operatorname{Tr}}

\def\ket#1{|#1\>}
\def\bra#1{\<#1|}

\newcommand{\hilb}[1]{\mathcal{#1}}
\newcommand{\td}[1]{\tilde{#1}}

\def\<{\langle}
\def\>{\rangle}

\newcommand{\mc}[1]{\mathcal{#1}}

\def\meb{\omega}

\begin{document}

\title{Generalized Hofmann quantum process fidelity bounds for quantum filters}

\author{Michal Sedl\'ak}
\affiliation{Department of Optics, Palack\'{y} University, 17. listopadu 1192/12, CZ-771 46 Olomouc, Czech Republic}
\affiliation{Institute of Physics,~Slovak Academy of Sciences,~D\'ubravsk\'a cesta 9,~84511 Bratislava,~Slovakia}

\author{Jarom\'{\i}r Fiur\'{a}\v{s}ek}
\affiliation{Department of Optics, Palack\'{y} University, 17. listopadu 1192/12, CZ-771 46 Olomouc, Czech Republic}

\begin{abstract}

We propose and investigate bounds on quantum process fidelity of quantum filters, i.e. probabilistic quantum operations represented by a single Kraus operator $K$.
These bounds generalize the Hofmann bounds on quantum process fidelity of unitary operations [H.F. Hofmann, Phys. Rev. Lett. \textbf{94}, 160504 (2005)],
and are based on probing the quantum filter by pure states forming two mutually unbiased bases. Determination of these bounds therefore requires much less measurements than full quantum process tomography.
We find that it is particularly suitable to construct one of the probe basis from the right eigenstates of $K$,
because in this case the bounds are tight in the sense that if the actual filter coincides with the ideal one then both the lower and upper bounds are equal to one.
We theoretically investigate application of these bounds to a two-qubit optical quantum filter formed by interference of two photons on a partially polarizing beam splitter.
For experimentally convenient choice of factorized input states and measurements we study the tightness of the bounds.
We show that more stringent bounds can be  obtained by more sophisticated processing of the data using convex optimization and we compare our methods for different choice of the input probe states.

\end{abstract}


\maketitle

\section{Introduction}

Characterization of quatum processes represents an indispensable tool for testing, optimization, and benchmarking of quantum information processing devices.
A complete characterization of a quantum device can be provided by quantum process tomography \cite{Poyatos97,Chuang97}, but this technique becomes very time consuming
with increasing complexity of the device, unless special procedure such as compressed sensing can be applied \cite{Gross10,Shabani11}, because the number
of parameters that need to be estimated scales exponentially with the number of qubits.

Instead of full quantum tomography we may just attempt to obtain an indication how close we are to the target operation, as quantified by the quantum process fidelity.
Monte Carlo sampling has been proposed for efficient estimation of fidelity of multiqubit states and operations with resources scaling polynomially with the estimation precision \cite{Flammia11,Silva11}.
Yet another experimentally appealing option is represented by the Hofmann bounds on quantum process fidelity \cite{Hofmann05}. In this approach, a lower and upper bound on fidelity with a unitary operation
is determined from measurements of average output state fidelities for input states forming  two mutually unbiased bases. This latter approach is particularly efficient for characterization of few-qubit operations
and it can be used e.g. for preliminary benchmarking of a quantum device during its design and optimization before a more complete characterization is carried out at the optimal operating point.
During recent years, the Hofmann bound  has been successfully utilized for experimental characterization of various two-qubit and three-qubit quantum gates \cite{Okamoto05,Bao07,Clark09,Gao10,Zhou11,Lanyon11,Micuda13,Micuda14}.

The Hofmann bound was designated to provide bounds on quantum process fidelity with a deterministic  unitary operation. Here we generalize this technique and propose and investigate Hofmann-like quantum process fidelity bounds
for special kind of probabilistic quantum operations called quantum filters. Quantum filters are completely positive trace decreasing maps that can be represented by a single Kraus operator $K$.
Quantum filters form an important tool in many branches of quantum information science and beyond and they find applications e.g. in quantum state engineering,
entanglement distillation \cite{Bennett96,Bennett96b,Kwiat01}, or quantum state discrimination \cite{Chefles00,Bergou04}.

Our derivation of the generalized Hofmann bounds for quantum filters is based on operator inequalities that are at the heart of the original Hofmann bound. In contrast to unitary operations, where measurement
of state fidelities for two complementary bases is sufficient, in case of quantum filters one generally needs to perform an additional set of measurements, which essentially characterizes the performance
of the filter  in a basis of its eigenstates. The number of measurements can be kept the same as for unitary operations provided that one of the two input bases is formed by the right eigenstates of $K$.
In this case it can be also proved that the resulting lower and upper bounds on quantum process fidelity are tight in the sense that for a perfect filter the bounds are always equal to $1$.
The probabilistic and non-unitary nature
of the quantum filters thus leads to a symmetry breaking and occurrence of a preferred set of probe input states. We explicitly consider two bases connected via Fourier transform
and also two $n$-qubit bases connected by Hadamard transform on each qubit.

As an illustration, we theoretically investigate characterization of a two-qubit optical quantum filter formed by interference of two photons on a partially polarizing beam splitter followed by post-selection
of detection of a single photon at each output port of the beam splitter. This filter is utilized in various linear optical quantum information processing devices such as linear optical quantum gates
\cite{Okamoto05,Kiesel05,Langford05,Kok07,Micuda13}.
We consider experimentally convenient choice of input probe states for which the required output state fidelities can be directly determined by product single-qubit measurements.
We show that as a consequence of this basis choice the resulting upper and lower bounds are not tight. We numerically find the ultimate upper and lower bounds for the same data using the semidefinite programming approach.
In this way we illustrate that for the considered quantum filter and the available data more stringent bounds can be obtained by more sophisticated processing of the data and we compare the two methods also
for different choice of basis states.

The rest of the paper is organized as follows. In section \ref{sec:boundsforfilters} we review the original Hofmann bound on fidelity of a deterministic process with a fixed unitary process
and we generalize this bound to quantum filters.
In section \ref{sec:ppbs} we study quantum filter formed by interference of two photons on a partially polarizing beam splitter and section \ref{sec:conclusions} contains the conclusions.
Finally, the Appendix contains a proof of an alternative lower bound for $n$-qubit filters and some technical derivations.

\section{Fidelity bounds for quantum filters}
\label{sec:boundsforfilters}

In general any quantum operation can be represented using Choi-Jamiolkowski isomorphism \cite{Jamiolkowski72,Choi75} by positive-semidefinite operators.
In this way a quantum filter $\mathcal{F}:\rho\mapsto K \rho K^\dagger$ is represented by an operator $\chi_\mathcal{F}=\ket{\omega_K} \bra{\omega_K}$, where
$\ket{\omega_K}=I\otimes K \ket{\omega}$,
\begin{equation}
\ket{\omega}=\frac{1}{\sqrt{d}}\sum_{j=1}^d \ket{e_j}\ket{e_j}
\end{equation}
denotes a maximally entangled state and $\{\ket{e_j}\}_{j=1}^d$ is an orthonormal basis of a $d$-dimensional Hilbert space $\hilb{H}_d$.
Fidelity between the actually implemented quantum operation  $\chi$ and the ideal quantum filter $\chi_\mathcal{F}$ can be defined via state fidelity between normalized Choi operators \cite{Horodecki99},
\begin{align}
F=\frac{\Tr(\chi\;\chi_\mc{F})}{\Tr(\chi)\Tr(\chi_\mc{F})}.
\end{align}
Our aim is to propose a procedure that would lower and upper bound this fidelity based on measured state fidelities of output states with respect to the ideal output for several input states.
To do this we first review the original Hofmann bound for deterministic operations in $d$ dimensions \cite{Hofmann05} and then we generalize it so that it will become applicable to non-unitary quantum filters.

Suppose $\{\ket{e_j}\}_{j=1}^d$, $\{\ket{f_k}\}_{k=1}^d$ are two orthonormal bases of $\hilb{H}_d$ that are mutually related by discrete Fourier transform,
i.e.
\begin{equation}
\ket{f_k}=\frac{1}{\sqrt{d}}\sum_{j=1}^d e^{i\frac{2\pi}{d}j k} \ket{e_j}.
\end{equation}
We denote by $\ket{e^{\mathrm{out}}_j}\equiv U \ket{e_j}$, $\ket{f^{\mathrm{out}}_k}\equiv U \ket{f_k}$
the ideal output states of a unitary transformation $U$. Density matrices  $\rho_j$ and $\xi_k$ of output states produced by the actually implemented operation $\chi$ from the input states $\ket{e_j}$, $\ket{f_k}$
can be expressed as
\begin{eqnarray}
\rho_j &=& d \Tr_{\rm in} (\chi \;\ket{e_j}\bra{e_j}^T\otimes I), \nonumber \\
\qquad \xi_k &=& d \Tr_{\rm in} (\chi \;\ket{f_k}\bra{f_k}^T\otimes I),
\label{chixidefinition}
\end{eqnarray}
where the transposition is taken with respect to the basis $\ket{e_j}$ used in the Choi-Jamiolkowski isomorphism.
The average output state fidelities for the two sets of probe states are defined as follows
\begin{eqnarray}
F_1&=&\frac{1}{d}\sum_{j=1}^{d} \bra{e^{\mathrm{out}}_j}\rho_j \ket{e^{\mathrm{out}}_j}, \nonumber \\
F_2&=&\frac{1}{d}\sum_{k=1}^{d} \bra{f^{\mathrm{out}}_k}\xi_k \ket{f^{\mathrm{out}}_k}.
\end{eqnarray}
The Hofman lower bound on quantum process fidelity \cite{Hofmann05}
\begin{align}
\label{eg:hfbound1}
F\geq F_1+F_2-1
\end{align}
can be proved as follows. For deterministic transformations $\Tr(\chi)=1$,  hence the bound (\ref{eg:hfbound1}) is equivalent to
\begin{align}
\label{eq:hfboundtes}
\Tr(\chi \;\ket{\omega_U}\bra{\omega_U})\geq & \sum_{j=1}^{d} \Tr (\chi \;\ket{e_j}\bra{e_j}^T\otimes \ket{e^{\mathrm{out}}_j}\bra{e^{\mathrm{out}}_j}) \nonumber \\
& +\sum_{k=1}^{d} \Tr (\chi \;\ket{f_k}\bra{f_k}^T\otimes \ket{f^{\mathrm{out}}_k}\bra{f^{\mathrm{out}}_k}) \nonumber \\
& - \Tr (\chi I\otimes I).
\end{align}
The validity of the inequality (\ref{eq:hfboundtes}) would be guaranteed by showing the positive-semidefiniteness of an operator $X=I\otimes U\;R\;I\otimes U^\dagger$, where
\begin{align}
\label{eq:hfopR}
R=&\ket{\omega}\bra{\omega}- \sum_{j=1}^{d} \ket{e_j}\bra{e_j}^T\otimes \ket{e_j}\bra{e_j} \nonumber \\
& -\sum_{k=1}^{d} \ket{f_k}\bra{f_k}^T\otimes \ket{f_k}\bra{f_k} + I\otimes I.
\end{align}
If $X\geq 0$ then $\Tr(\chi X)\geq 0$ due to positivity of $\chi\geq 0$, which implies the inequality (\ref{eq:hfboundtes}).
As we show in the appendix \ref{sec:positivityofR} the operator $R$ can be rewritten (in the term to term fashion) as
\begin{align}
\label{eq:hfRdiag}
R=&\ket{\meb_{11}} \bra{\meb_{11}} -\sum_{j=1}^d \ket{\meb_{j1}} \bra{\meb_{j1}}-\sum_{k=1}^d \ket{\meb_{1k}} \bra{\meb_{1k}} \nonumber \\
&+ \sum_{j,k=1}^d \ket{\meb_{jk}} \bra{\meb_{jk}},
\end{align}
where $\{\ket{\meb_{jk}}\}$ is an orthonormal basis of maximally entangled states in $d$ dimensions.

From the above equation it is clear that $R\geq 0$, which implies $X\geq 0$ and this proves the original Hofmann bound.
In a similar fashion, the upper bounds on quantum process fidelity $F\leq F_1$, $F\leq F_2$ can be derived from the following two operator inequalities
\begin{equation}
\label{upperinequalities}
\ket{\meb_{11}} \bra{\meb_{11}}\leq \sum_{j=1}^d \ket{\meb_{j1}} \bra{\meb_{j1}}, \quad \ket{\meb_{11}} \bra{\meb_{11}}\leq \sum_{k=1}^d \ket{\meb_{1k}} \bra{\meb_{1k}}.
\end{equation}

Next, we will use the operator $R$ to derive lower and upper bound on the fidelity of quantum filters. Let us multiply $R$
by $I\otimes K$ from the left and by $I\otimes K^\dagger$ from the right. Using  Eq. (\ref{eq:hfopR}) we obtain
\begin{align}
\label{eq:KRK}
&\ket{\omega_K}\bra{\omega_K}- \sum_{j=1}^{d} \ket{e_j}\bra{e_j}^T\otimes K\ket{e_j}\bra{e_j}K^\dagger \nonumber \\
& -\sum_{k=1}^{d} \ket{f_k}\bra{f_k}^T\otimes K\ket{f_k}\bra{f_k}K^\dagger + I\otimes K K^\dagger\geq 0,
\end{align}
where the inequality follows from $R\geq 0$. By taking the trace with $\chi$ Eq. (\ref{eq:KRK}) can be rewritten as
\begin{align}
\label{eq:KRKineq1}
\Tr(&\chi \;\ket{\omega_K}\bra{\omega_K})- \frac{1}{d}\sum_{j=1}^{d} \Tr(K\ket{e_j}\bra{e_j}K^\dagger\; \rho_j) \nonumber \\
&- \frac{1}{d}\sum_{k=1}^{d} \Tr(K\ket{f_k}\bra{f_k}K^\dagger\; \xi_k) 
+\Tr(K K^\dagger\; \Omega) \geq 0,
\end{align}
where $\rho_j$ and $\xi_k$ defined in Eq.~(\ref{chixidefinition}) are the unnormalized output states of a probabilistic operation $\chi$ corresponding to pure input states $\ket{e_j}$ and $\ket{f_k}$, respectively, and
$\Omega=d \Tr_{\rm in} (\chi \;(\frac{1}{d}I)^T\otimes I)$ is the unnormalized output state for a maximally mixed input state. 
To obtain  a lower bound on fidelity of a quantum filter $K$,  we divide the inequality (\ref{eq:KRKineq1}) by $\Tr(\chi)\Tr(\chi_\mc{F})$ and rewrite the resulting expression such that it contains normalized overlaps,
\begin{align}
\label{eq:KRKineq2}
F\geq &\sum_{j=1}^{d} p_j \bra{\td{e}_j}\td{\rho}_j\ket{\td{e}_j} 
+ \sum_{k=1}^{d} q_k \bra{\td{f}_k}\td{\xi}_k\ket{\td{f}_k} 
-\Delta\, \Tr(K K^\dagger\; \td{\Omega}).
\end{align}
Here  $\td{\rho}_j=\rho_j/ \Tr(\rho_j)$,  $\td{\xi}_k=\xi_k / \Tr(\xi_k)$,  and
 $\td{\Omega}=\Omega / \Tr(\Omega)$,
are normalized output states of $\chi$ and
\begin{equation}
\ket{\td{e}_j}=\frac{K\ket{e_j}}{\sqrt{\bra{e_j}K^\dagger K \ket{e_j}}}, \quad
\ket{\td{f}_k}=\frac{K\ket{f_k}}{\sqrt{\bra{f_k}K^\dagger K \ket{f_k}}},
\label{efnormalized}
\end{equation}
denote the normalized output states of the ideal filter $K$. The weight factors read
\begin{align}
\label{eq:KRKineq2def}
p_j&=\Delta P_j \bra{e_j}K^\dagger K \ket{e_j}, \quad 
q_k=\Delta Q_k \bra{f_k}K^\dagger K \ket{f_k},\nonumber \\
\end{align}
 where $\Delta=d/\Tr(K^\dagger K)=1/\Tr(\chi_\mc{F})$, and  $P_j=\frac{\Tr(\rho_j)}{d\Tr(\chi)}$ and $Q_k=\frac{\Tr(\xi_k)}{d\Tr(\chi)}$ denote
the relative success probabilities of operation $\chi$ for input basis states $\ket{e_j}$ and $\ket{f_k}$, respectively. These relative probabilities satisfy $\sum_j P_j=\sum_k Q_k=1$.

Formula (\ref{eq:KRKineq2}) generalizes the Hofmann lower bound (\ref{eg:hfbound1}) to quantum filters and represents one of our main results. We can see that
the fidelity of a quantum filter is lower bounded by an expression which contains two weighted sums of the output state fidelities $\bra{\td{e}_j}\td{\rho}_j\ket{\td{e}_j}$
and  $\bra{\td{f}_k}\td{\xi}_k\ket{\td{f}_k}$, which generalizes the average state fidelities $F_1$ and $F_2$ appearing in the original Hofmann bound.
The last term on the right-hand side of inequality (\ref{eq:KRKineq2}) provides a generalization of the factor $-1$ to quantum filters and depends both on the ideal filter $K$ and also on the actual operation
$\chi$ through $\tilde{\Omega}$. It follows from Eq. (\ref{eq:KRKineq1}) that the summation in Eq. (\ref{eq:KRKineq2}) should be performed only over those terms for which the overlap
$\bra{e_j}K^\dagger K \ket{e_j}$ or $\bra{f_k}K^\dagger K \ket{f_k}$ is nonzero. This also ensures that the normalized output states (\ref{efnormalized}) are well defined.

A most straightforward way to experimentally determine the output state fidelities and relative success probabilities $P_j$ and $Q_k$ consists in measuring the output state $\rho_j$ ($\xi_k$) in a basis including the
corresponding output state $\ket{\tilde{e}_j}$ ($\ket{\tilde{f}_k}$) produced by an ideal filter. For quantum filters $\{\ket{\tilde{e}_j}\}_{j=1}^d$ and $\{\ket{\tilde{f}_k}\}_{k=1}^d$ generally do not form a basis
which means that a separate measurement basis must be set for each probe state. Besides testing the unknown quantum transformation with $2d$
input states  $\{\ket{e_j}\}_{j=1}^d$, $\{\ket{f_k}\}_{k=1}^d$, we also need to determine the term $\Tr(K K^\dagger \td{\Omega})$ by some measurements.
To construct a suitable measurement, consider a singular value decomposition of $K$,
\begin{align}
\label{eq:kspectral}
K =\sum_{l=1}^d \sqrt{\lambda_l} \ket{v_l}\bra{w_l},
\end{align}
where the left and right eigenvectors $\ket{v_l}$, $\ket{w_l}$ form two orthonormal bases and the non-negative singular values were chosen in the form $\sqrt{\lambda_l}$ to simplify subsequent formulas.
As a consequence the positive-semidefinite operators $K^\dagger K$, $K K^\dagger$ have the following spectral decompositions
\begin{align}
\label{eq:kkdspectral}
K^\dagger K=\sum_{l=1}^d \lambda_l \ket{w_l}\bra{w_l}, \qquad K K^\dagger=\sum_{l=1}^d \lambda_l \ket{v_l}\bra{v_l}.
\end{align}
In principle, we can determine $\Tr(K K^\dagger \td{\Omega})$ from suitable measurements on any $d$ input states forming an orthonormal basis (e.g. vectors $\ket{u_j}$).
Let $\zeta_j=d \Tr_{\mathrm{in}}(\chi \;\ket{u_j}\bra{u_j}^T\otimes I)$ and $\td{\zeta}_j=\zeta_j/\Tr(\zeta_j)$ denote the unnormalized and normalized output state corresponding to the input state $\ket{u_j}$,
and similarly as before we define the relative success probability for this input state as $R_j=\frac{\Tr(\zeta_j)}{d\Tr(\chi)}$.
The term $\Tr(K K^\dagger\; \td{\Omega})$ can then be expressed as
\begin{align}
\label{eq:kkomega}
\Tr(K K^\dagger\; \td{\Omega})
=\sum_{j,l=1}^d \lambda_l R_j\bra{v_l}\td{\zeta}_j\ket{v_l}.
\end{align}
The relative success probabilities $R_j$ as well as the overlaps of output states $\td{\zeta}_j$ with  $\ket{v_l}$ can be determined by measuring each output state $\td{\zeta}_j$
in the basis formed by the eigenstates $\ket{v_l}$.

At this point it is useful to realize that 
the experimental effort can be kept the same as for $K$ being unitary at the price of a suitable choice of basis 
$\ket{e_j}$ and consequently $\ket{f_k}$. Thus, if we choose $\ket{e_j}=\ket{w_j}$ and also $\ket{u_j}=\ket{w_j}$ then
\begin{align}
\label{eq:etd}
\ket{\td{e}_j}&=\frac{K\ket{w_j}}{\sqrt{\bra{w_j}K^\dagger K \ket{w_j}}}= \ket{v_j}
\end{align}
and the data for input states $\ket{e_j}$ measured after the filter in basis $\ket{\td{e}_j}$ can be used to determine the last term of Eq. (\ref{eq:KRKineq2}), i.e. $\td{\zeta}_j=\td{\rho}_j$, $R_j=P_j$ $\forall j$.
After some algebra we find that
the lower bound now equals to
\begin{align}
\label{eq:KRKineqfin}
F\geq  \sum_{k=1}^{d}  Q_k \bra{\td{f}_k}\td{\xi}_k\ket{\td{f}_k}-\sum_{j,l=1}^d \frac{\lambda_l}{\overline{\lambda}}(1-\delta_{jl}) P_j\bra{\td{e}_l}\td{\rho}_j\ket{\td{e}_l},
\end{align}
where $\overline{\lambda}\equiv (\sum_{j=1}^d \lambda_j)/d$ and we used the identity $\bra{f_k}K^\dagger K \ket{f_k}=\overline{\lambda}$, which holds since the two bases $\ket{e_j}$ and
$\ket{f_k}$ are related by quantum Fourier transform.

An important property of the lower bound is its tightness. Especially, if the implemented transformation is the desired one, then fidelity $F=1$ and we want our lower bound to attain the value $1$ as well.
If the implementation of the filter is perfect, then $\bra{\td{f}_k}\td{\xi}_k\ket{\td{f}_k}=1$ and $\bra{\td{e}_l}\td{\rho}_j\ket{\td{e}_l}=\delta_{jk}$. If we insert these expressions into Eq.~(\ref{eq:KRKineqfin})
then we get $F\geq 1$ which confirms that the lower bound (\ref{eq:KRKineqfin}) is tight for any quantum filter $K$. Note that this tightness is achieved due to the special choice of the probe states, where
 $\ket{e_j}$ coincide with the right eigenvectors of $K$. For other choices of the probe states the lower bound (\ref{eq:KRKineq2}) is generally not tight and can be strictly lower than $1$ even for
a perfect filter. This should be contrasted with the original Hofmann bound (\ref{eg:hfbound1}) which always attains value $1$ if the  target unitary $U$ is implemented perfectly,
irrespective of the choice of the two probe bases. The nonunitarity of the filter $K$ thus leads to a symmetry breaking and emergence of preferred states suitable for benchmarking of the filter.

In a very similar fashion as above also a pair of upper bounds can be derived
\begin{align}
\label{eq:KRKinequp0}
F\leq \sum_{j=1}^{d} p_j \bra{\td{e}_j}\td{\rho}_j\ket{\td{e}_j}, \qquad
F\leq \sum_{k=1}^{d} q_k \bra{\td{f}_k}\td{\xi}_k\ket{\td{f}_k},
\end{align}
if we start from inequalities
\begin{align}
&\ket{\omega}\bra{\omega}\leq\sum_{j=1}^{d} \ket{e_j}\bra{e_j}^T\otimes \ket{e_j}\bra{e_j}, \nonumber \\
&\ket{\omega}\bra{\omega}\leq \sum_{k=1}^{d} \ket{f_k}\bra{f_k}^T\otimes \ket{f_k}\bra{f_k},
 \end{align}
 that are equivalent to the inequalities (\ref{upperinequalities}). For the special choice of probe states $\ket{e_j}=\ket{w_j}$ the upper bounds simplify to
\begin{align}
\label{eq:KRKinequp}
F\leq \sum_{j=1}^{d} \frac{\lambda_j}{\overline{\lambda}}  P_j \bra{v_j}\td{\rho}_j\ket{v_j}, \qquad
F\leq \sum_{k=1}^{d} Q_k \bra{\td{f}_k}\td{\xi}_k\ket{\td{f}_k}.
\end{align}
Similarly as the lower bound (\ref{eq:KRKineqfin}), also the upper bounds (\ref{eq:KRKinequp}) are tight in the sense that they yield $F\leq 1$ if the filter is implemented perfectly.

From the application point of view, the $n$-qubit systems whose Hilbert space is endowed with  tensor product structure and $d=2^n$ are particularly relevant. In this case a natural choice of $\ket{e_j}$
could be the computational basis formed by tensor products of single-qubit states $\ket{0}$ and $\ket{1}$. By its construction, discrete quantum Fourier transform of the $n$-qubit computational basis states
leads to product $n$-qubit states $\ket{f_k}$. This is good for experiments, since preparation of product states is often much simpler than preparation of entangled states. Unfortunately,
for most quantum filters at least some (ideal) output quantum states for the above inputs are entangled. For that reason it might be useful to study also other pairs of input bases,
which could be tested more easily. One such combination can be the computational basis and the Hadamard basis formed by tensor products of states $\ket{\pm}=(\ket{0}\pm\ket{1})/\sqrt{2}$.
As we show in the Appendix~\ref{sec:altlb} the above derived lower and upper bounds hold also for this latter setting.

\section{Two-qubit quantum filters}
\label{sec:ppbs}

The Hofmann bound (\ref{eg:hfbound1}) proved particularly useful for characterization of various linear optical quantum gates \cite{Okamoto05,Bao07,Clark09,Gao10,Zhou11,Micuda13,Micuda14}.
As a case study, we therefore investigate here a characterization  of a quantum filter acting on polarization state
of two photons. The filtering is realized by interference of the photons on a partially polarizing beam splitter (PPBS)
followed by postselection of events when a single photon is observed at each output port of the beam splitter.
In the basis of vertical and horizontal polarizations, the resulting two-qubit quantum filter reads
\begin{align}
K = \left(
\begin{array}{cccc}
t_H^2-r_H^2 & 0 & 0 & 0 \\
0 & t_H t_V & -r_H r_V & 0 \\
0 & -r_H r_V & t_H t_V & 0 \\
0 & 0 & 0 & t_V^2-r_V^2
\end{array}
\right),
\end{align}
where $t_H$, $t_V$ and $r_H$, $r_V$ denote the amplitude transmittances and reflectances of PPBS for horizontal and vertical polarizations, respectively.
We assume that the transmittances and reflectances are real and that the beam splitter is lossless, hence $t_j^2+r_j^2=1$.

\begin{figure}[!t!]
\includegraphics[width=0.95\linewidth]{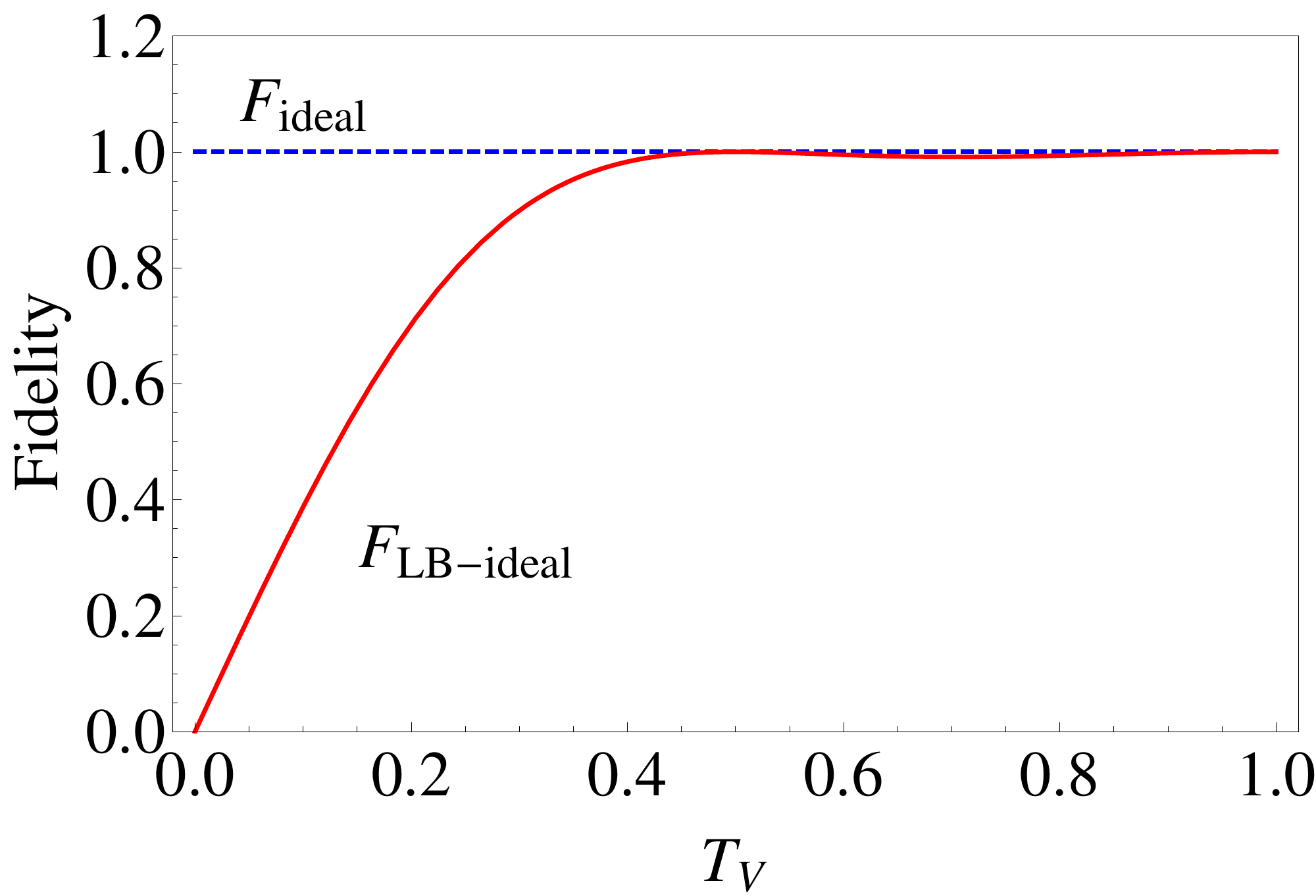}
\caption{A lower bound (\ref{eq:KRKineq2}) on fidelity $F$ of a two-qubit quantum filter $K$ is plotted in dependence on the intensity transmittance $T_V=t_V^2$ of the partially polarizing beam splitter.
Ideal implementation of the filter is assumed, hence the true fidelity $F=1$ and is depicted by the blue dashed line.
The gap between the two lines illustrates the tightness of the bound.}
\label{fig:tightness1}
\end{figure}

We will investigate further only the case $t_H=1$, since this element is often used in optical quantum information processing experiments and its
fidelity should be assessed. In this case the filter becomes diagonal, $K=\mathrm{diag}\{1,t_V,t_V, 2t_V^2-1\}$.
Choosing $\{\ket{e_j}\}_{j=1}^4$ as the computational basis in which the filter is diagonal would for probe states $\ket{f_k}$ lead to measurements on output states in an entangled basis,
which is problematic in many optical experimental setups. Instead, we will introduce alternative probe states that require just preparations and measurements in product bases.
 The idea is to employ the following pair of bases,
\begin{align}
\label{probestates}
\ket{e_1}=\ket{0}\ket{+}, \qquad  \ket{f_1}=\ket{+}\ket{0}, \nonumber \\
\ket{e_2}=\ket{0}\ket{-}, \qquad  \ket{f_2}=\ket{+}\ket{1},  \\
\ket{e_3}=\ket{1}\ket{+}, \qquad \ket{f_3}=\ket{-}\ket{0}, \nonumber \\
\ket{e_4}=\ket{1}\ket{-}, \qquad \ket{f_4}=\ket{-}\ket{1}. \nonumber
\end{align}
The choice of these probe states is motivated by previous experiments, where bounds on fidelity of a quantum controlled-NOT gate and controlled-Z gate
were determined \cite{Okamoto05,Bao07,Clark09,Gao10,Micuda14}. The two bases (\ref{probestates}) are related via a Hadamard transform on each qubit,
$\ket{f_j}=H\otimes H\ket{e_j}$, and this relation together with the factorized form of the basis states $\ket{e_j}$ and $\ket{f_k}$
ensures that the lower and upper fidelity bounds (\ref{eq:KRKineq2}) and (\ref{eq:KRKinequp0}) are applicable, c.f. also Appendix B. The practical advantage of the probe states (\ref{probestates})
is that the filter $K$ maps them on product states,
\begin{align}
&\ket{\td{e}_1}=\ket{0}\ket{a_+}, \quad \quad \ket{\td{f}_1}=\ket{a_+}\ket{0}, \nonumber \\
&\ket{\td{e}_2}=\ket{0}\ket{a_-}, \quad \quad \ket{\td{f}_2}=\ket{b_+}\ket{1}, \nonumber \\
&\ket{\td{e}_3}=\ket{1}\ket{b_+}, \quad \quad \ket{\td{f}_3}=\ket{a_-}\ket{0}, \nonumber \\
&\ket{\td{e}_4}=\ket{1}\ket{b_-}, \quad \quad \ket{\td{f}_4}=\ket{b_-}\ket{1}, \nonumber 
\end{align}
where
\begin{align}
\ket{a_{\pm}}&=\frac{1}{\sqrt{1+t_V^2}}\ket{0}\pm \frac{t_V}{\sqrt{1+t_V^2}}\ket{1}, \nonumber \\
\ket{b_{\pm}}&=\frac{t_V}{\sqrt{t_V^2+(2t_V^2-1)^2}}\ket{0}\pm \frac{2t_V^2-1}{\sqrt{t_V^2+(2t_V^2-1)^2}}\ket{1}. \nonumber 
\end{align}

\begin{figure}[!t!]
\includegraphics[width=\linewidth]{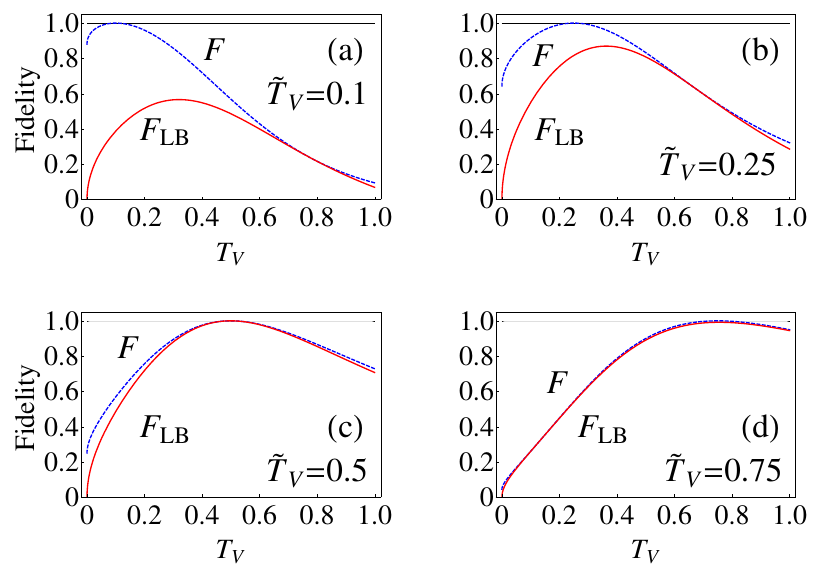}
\caption{The red solid lines show the dependence of the lower bound and fidelity of a quantum filter $K$ on the actual intensity transmittance $T_V=t_V^2$ of the PPBS
 for four different target transmittances $0.1$ (a), $0.25$ (b), $0.5$ (c), and $0.75$ (d). The true fidelity of the filter is depicted by blue dashed lines.}
\label{fig:diff1}
\end{figure}

Since the filter $K$ is not diagonal in any of the two probe bases (\ref{probestates}), additional measurements are required to estimate the term $\Tr(KK^\dagger \tilde{\Omega})$.
 A natural choice is to employ the computational basis states as additional probes $\ket{u_j}$ and measure the resulting output states in the computational basis. 
 Mathematically, we have
 $\ket{u_j}=\ket{w_j}=\ket{v_j}$ and
\begin{align}
\ket{v_1}=\ket{00}, \quad
\ket{v_2}=\ket{01}, \quad
\ket{v_3}=\ket{10}, \quad
\ket{v_4}=\ket{11}.
\end{align}
The above outlined procedure to test the unknown quantum filter requires $3d=12$ product input states and each output state needs to be measured in a single product basis.
For comparison, full quantum process tomography of a two-qubit quantum filter \cite{Mitchell03} would typically involve about $144$ different combinations of input states and output measurements.
Thus, characterization of the filter via the generalized Hofmann bounds
requires much less resources.

\begin{figure*}[!t!]
\centerline{\includegraphics[width=\linewidth]{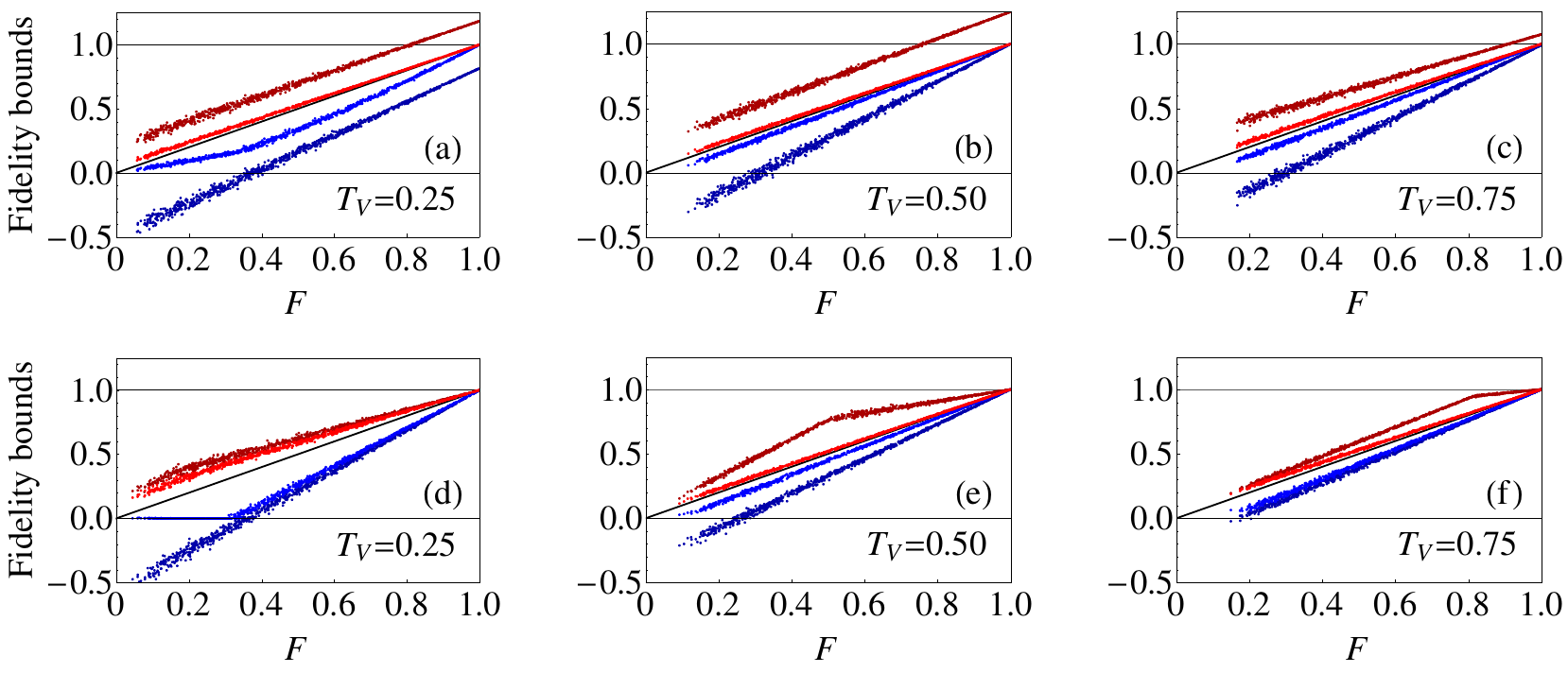}}
\caption{Graphs of upper and lower bounds on the fidelity of two-qubit quantum filters. Upper and lower most dots correspond to analytical bounds
for randomly generated filters constructed as a mixture of a fixed ideal filter with transmittance $t_V$ specified in each panel and a randomly chosen filter.
The dots just above/below the solid line correspond to the bounds obtained from the same data, but using semidefinite programming approach.
The results in the upper row correspond to probing with states (\ref{probestates}) for which measurements in product basis are sufficient.
For comparison, the three graphs in the lower row were obtained for a different set of probe states with
$\ket{e_j}$ chosen to be the computational basis where the ideal filter $K$ is diagonal. This latter choice of probe states requires that some of the output states are measured in entangled basis.}
\label{fig:numresults}
\end{figure*}

Let us now discuss the tightness of the lower bound (\ref{eq:KRKineq2}) for our choice of the filter and the probe states. A detailed analysis reveals that in the present case this bound
is tight for the ideal quantum filter, i.e. equal to one, only if $T_V=\frac{1}{2}$ or $T_V=1$, where $T_V=t_V^2$. Otherwise there is a gap, which is smaller than $1\%$ if the desired PPBS has $T_V>\frac{1}{2}$,
see Fig.~\ref{fig:tightness1}. We also considered the situation where the actual value of the transmittance $T_V$ differs from the desired one. The dependence of the
lower bound on the actual value of $T_V$ is plotted in  Fig.~\ref{fig:diff1} for four different desired values of $T_V$. 
The results  indicate that the proposed lower bound works well for $T_V>\frac{1}{2}$. However, for $T_V<\frac{1}{2}$ the bound quickly becomes too loose to be useful.

This opens the question what are the best upper and lower bounds on fidelity given certain data from experiment. Such a question can be precisely answered by semidefinite programming \cite{Bartuskova06},
since the fidelity and the constraints given by the measured data are linear in $A=\frac{\chi}{\Tr(\chi)}$. Thus, we might write
\begin{align}
F_{\mathrm{UB}}=\max_{\substack{A\geq 0,\, \Tr(A)=1 \\ \forall k \Tr(A M_k)=r_k}} \Delta \Tr(A \ket{\omega_K} \bra{\omega_K}) \nonumber \\ 
F_{\mathrm{LB}}=\min_{\substack{A\geq 0,\, \Tr(A)=1 \\ \forall k \Tr(A M_k)=r_k}} \Delta \Tr(A \ket{\omega_K} \bra{\omega_K}) 
\label{sdp}
\end{align}
where the matrices $M_k$ and parameters $r_k$ capture the linear constraints provided by the data. To make our formulation of these constraints sufficiently general,
consider a set of input probe states $\{\ket{m_j}\}_{j=1}^d$ forming a basis,
where each output state is measured in a generally different basis $\{\ket{n_{jk}}\}_{k=1}^d$. Let $f_{jk}$ denote the measured frequencies which sample the theoretical probabilities
\begin{equation}
p_{jk}=d \Tr(\chi \; \ket{m_j}\bra{m_j}^{T}\otimes \ket{n_{jk}}\bra{n_{jk}}).
\end{equation}
Using the identity  $\sum_{j,k=1}^d p_{jk}=d\Tr(\chi)$ we can express the set of constraints on $A$ imposed by the data $f_{jk}$ as follows,
\begin{equation}
 \Tr(A \; \ket{m_j}\bra{m_j}^{T}\otimes \ket{n_{jk}}\bra{n_{jk}}) = \frac{f_{jk}}{\sum_{l,m=1}^d f_{lm}}.
\end{equation}
 In our present case, each  probe basis $\ket{e_j}$, $\ket{f_k}$, and $\ket{v_l}$ provides a set of $15$ linearly independent constraints.

 To test the performance of this approach, we used it to find the best upper and lower bounds from the data that we generated for $1000$ randomly chosen quantum operations $\chi$ that were constructed as a random mixture
 of an ideal quantum filter $K$ and some other randomly chosen filter $K'$. The  optimization  (\ref{sdp}) was performed numerically with the use 
  of CVX, a package for specifying and solving convex programs \cite{cvx1,cvx2}.
 For a given random quantum filter $\chi$ we also compute the analytical lower and upper fidelity bounds according to our procedure and the true fidelity.
 In figure \ref{fig:numresults} we plot the obtained results as a function of the true fidelity, hence the graph of the true fidelity forms a straight line with a unit slope. For comparison, we consider probing with
 two sets of input states. The first set is specified by Eq. (\ref{probestates}) and requires only product measurements on output states, but the analytical bounds are not tight. The second set of probe states
 is constructed such that it leads to a tight analytical bounds (\ref{eq:KRKineqfin}) and (\ref{eq:KRKinequp}) at the price of the requirement of measurements in entangled basis for some of the output states.
 Interestingly, the approach based on semidefinite programming provides in both cases tight bounds, i.e. for ideal implementation of the filter the upper and lower bounds coincide and $F_{\mathrm{LB}}=F_{\mathrm{UB}}=1$.

\section{Conclusions}
\label{sec:conclusions}

In summary, we designed and analyzed bounds on quantum process fidelity $F$ of a specific type of nondeterministic operations called quantum filters. These operations
are mathematically characterized as completely positive maps, which can be expressed by a single Kraus operator $K$. Operationally they correspond to operations
which succeed only with a limited probability that depends on the input state and they map any pure state again to pure state.
The proposed bounds represent a generalization of the original Hofmann bounds on fidelity of  unitary transformations \cite{Hofmann05}.
For quantum filters, the average state fidelities are replaced with specific weighted averages, and the lower bound contains also an additional
 term that depends both on the desired and the actual operation. As a consequence, in addition to
determination of relative success rates and output state fidelities for two sets of input basis states, further
measurements are generally needed.
 Nevertheless, we show that the number of input states and measurements can be kept the same as for unitary operations if one of the two input bases is formed by the right eigenstates of $K$.
An important property of any bound is its tightness. In particular, for quantum process fidelity bounds we would like to have both the upper and lower bounds equal to
one if the actual and the desired quantum transformation coincide, because in that case the fidelity is $F=1$. We demonstrated that our bound is tight if one set of probe states
 is formed by right eigenstates of $K$ and the other by their quantum Fourier transform or by their Hadamard transform.
 The proposed bounds extend the toolbox of efficient methods of characterization of quantum operations \cite{Gross10,Shabani11,Flammia11,Silva11,Hofmann05,Emerson07,Dankert09,Reich13}
 and provide a method for quick checking of quality of quantum filters before  their more comprehensive characterization, e.g. by quantum process tomography.

As an illustration, we have theoretically investigated application of the proposed fidelity bounds to characterization of a specific two-qubit linear-optical quantum filter.
This filter is implemented by two-photon interference on a partially polarizing beam splitter followed by conditioning on emergence of a single photon at each output port of the beam splitter,
and it is often utilized in optical quantum information processing with polarization encoded qubits.
We consider experimentally convenient choice of product input probe states for which the required output state fidelities can be directly determined by product single-qubit measurements.
It turns out that the price to pay for this experimental convenience is that the resulting bounds are generally not tight.
We compare our analytical bounds with ultimate lower and upper bounds
that can be obtained from a given experimental data with the help of convex optimization. The experimental data represent a set of linear constraints and we numerically
solve a so-called semidefinite program that among all quantum operations satisfying given linear constraints finds an operation with minimum and maximum overlap with the
target quantum filter $K$. We observe that the ultimate lower and upper fidelity bounds obtained in this way are tight, i.e. equal to one for perfect filters, even if the analytical ones are not.
In this way we illustrate that for the considered quantum filter and the available data more stringent bounds can be obtained
by more sophisticated data processing.

\begin{acknowledgments}
J.F. acknowledges support by the Czech Science Foundation (Project No. 13-20319S).
 M.S. acknowledges support by the European Social Fund and the state budget of the Czech Republic under Operational Program Education for Competitiveness
 (Project No. CZ.1.07/2.3.00/30.0004).
\end{acknowledgments}

\appendix

\section{Positivity of operator $R$}
\label{sec:positivityofR}
Let us define an orthonormal basis $\{\ket{\meb_{jk}}\}_{j,k=1}^d$ of maximally entangled states via the action of a pair of operators
\begin{align}
\label{eq:defZW}
Z=\sum_j \ket{e_{j\oplus 1}}\bra{e_j}, \quad
W=\sum_j e^{i \frac{2\pi j}{d}} \ket{e_j}\bra{e_j}, 
\end{align}
on the state $\ket{\omega}$ as
\begin{align}
\label{eq:defmeb}
\ket{\meb_{jk}}=Z^{j-1}W^{k-1}\otimes I \;\ket{\omega}.
\end{align}
By definition $\ket{\meb_{11}}=\ket{\omega}$ and
\begin{align}
\sum_{j=1}^{d} \ket{\meb_{1j}}\bra{\meb_{1j}}&=\frac{1}{d}\sum_{j,k,l=1}^d  e^{i \frac{2\pi j(k-l)}{d}}
\ket{e_k}\ket{e_k} \bra{e_l}\bra{e_l}\nonumber \\
&= \sum_{j=1}^{d} \ket{e_j}\bra{e_j}^T\otimes \ket{e_j}\bra{e_j},
\end{align}
because here the transposition is taken with respect to the basis $\ket{e_j}$.
Similarly,
\begin{align}
\label{eq:summebj0}
\sum_{j=1}^{d} \ket{\meb_{j1}}\bra{\meb_{j1}}&=\frac{1}{d}\sum_{j,k,l=1}^d
\ket{e_{k\oplus j}}\ket{e_k} \bra{e_{l\oplus j}}\bra{e_l}.
\end{align}
On the other hand,
\begin{align}
\sum_{k=1}^{d} &\ket{f_k}\bra{f_k}^T \otimes \ket{f_k}\bra{f_k}= \nonumber \\
&=\frac{1}{d}\sum_{j,k,l,m,n=1}^{d} e^{i \frac{2\pi k(l-j+m-n)}{d}}
\ket{e_j}\ket{e_l}\bra{e_m}\bra{e_n} \nonumber \\
&=\sum_{\substack{j,k,l,m,n \\ j-l=m-n}} \ket{e_j}\ket{e_l}\bra{e_m}\bra{e_n},
\end{align}
which is clearly equivalent to (\ref{eq:summebj0}). Thus the operator $R$ defined by Eq.~(\ref{eq:hfopR}) can be expressed as
\begin{align}
R=&\ket{\meb_{11}} \bra{\meb_{11}}-\sum_{j=1}^d \ket{\meb_{j1}} \bra{\meb_{j1}}-\sum_{k=1}^d \ket{\meb_{1k}} \bra{\meb_{1k}} \nonumber \\
&+ \sum_{j,k=1}^d \ket{\meb_{jk}} \bra{\meb_{jk}},
\end{align}
where we used the identity $I\otimes I= \sum_{j,k=1}^d \ket{\meb_{jk}} \bra{\meb_{jk}}$.
Since we managed to rewrite $R$ as a sum of projectors, $R=\sum_{j,k=2}^d \ket{\meb_{jk}} \bra{\meb_{jk}}$, this proves that $R\geq 0$.

\section{Alternative lower bound for $n$-qubit filters}
\label{sec:altlb}

Our goal is to prove positivity of operator $R$ defined in Eq. (\ref{eq:hfopR}) for a different pair of orthonormal bases $\{\ket{e_j}\}_{j=1}^d$, $\{\ket{f_k}\}_{k=1}^d$.
This would allow us to exactly repeat the same steps as in the main text and thus we could use all the derived lower and upper bounds, but for a different choice of $\ket{e_j}$, $\ket{f_k}$.
Specifically, we shall consider systems of $n$ qubits, hence $d=2^n$.
Let $\{\ket{e_j}\}_{j=1}^d$ be the computational basis, where $\ket{e_j}=\ket{j_1}\ket{j_2}\ldots\ket{j_n}$ and $j_m$ is the $m$-th digit in the binary representation of number $j-1$.
Similarly, let $\{\ket{f_k}\}_{k=1}^d$ be the computational basis transformed by the Hadamard transform (acting as $H\ket{j_m}=(\ket{0}+(-1)^{j_m}\ket{1})/\sqrt{2}$) on every qubit,
$\ket{f_k}=H\ket{k_1}\otimes H\ket{k_2}\ldots \otimes H\ket{k_n}$.
Operator $R$ is acting on $2n$ qubits, which are ordered as $n$ qubits related to the input state tensored with another $n$ qubits related to the output state of the quantum filter.
It is useful to divide the $2n$-qubit Hilbert space on which operator $R$ acts into two-qubit subsystems
formed by the $m$-th qubit of the input and the $m$-th qubit of the output. We introduce a unitary operator $W$ which groups together the $m$-th
input and output qubits \cite{Micuda13},
\begin{equation}
W|j_1\ldots,j_n\rangle|k_1,\ldots,k_n\rangle=|j_1,k_1\rangle\ldots|j_n,k_n\rangle.
\end{equation}
In this way the maximally entangled state can be written as
 $W\ket{\omega}=\ket{\Phi^{+}}_1 \cdots \ket{\Phi^{+}}_n$, where $\ket{\Phi^{\pm}}=\frac{1}{\sqrt{2}}(\ket{00}\pm\ket{11})$
 are the Bell states and the subscripts indicate the two-qubit subsystems. It is not difficult to show that the summations over all projectors
 $\ket{e_j}\bra{e_j}^T\otimes \ket{e_j}\bra{e_j}$ and $\ket{f_k}\bra{f_k}^T\otimes \ket{f_k}\bra{f_k}$ in Eq. (\ref{eq:hfopR})
 factorize into products of $n$ summations over two-qubit subsystems consisting of a single input and output qubit.
 The summations over the two-qubit subspaces can be performed with the use of the following identities
\begin{align}
\ket{00}\bra{00}+\ket{11}\bra{11}&= \Phi^{+} + \Phi^{-},  \nonumber \\
\ket{++}\bra{++}+\ket{--}\bra{--}&= \Phi^{+} + \Psi^{+},
\label{psisums}
\end{align}
where $\ket{\Psi^{\pm}}=\frac{1}{\sqrt{2}}(\ket{01}\pm\ket{10})$ are the other two Bell states and we used the notation $\Phi^{+}\equiv \ket{\Phi^{+}}\bra{\Phi^{+}}$.
 The identities (\ref{psisums}) allow us to rewrite  the operator $R$ as
\begin{align}
\label{psumid}
 W R\, W^\dagger =&(\Phi^{+})^{\otimes n} - (\Phi^{+} + \Phi^{-})^{\otimes n} - (\Phi^{+} + \Psi^{+})^{\otimes n} \nonumber \\
 &+ (\Phi^{+} + \Phi^{-}+ \Psi^{+}+ \Psi^{-})^{\otimes n},  \nonumber \\
\end{align}
where we used the four Bell states to express the identity on the two qubit Hilbert space as $\mathbb{I}=\Phi^{+} + \Phi^{-} + \Psi^{+} + \Psi^{-}$.
Since $W$ is unitary, operator $WR\,W^\dagger$ has the same eigenvalues as $R$. Moreover,
$WR\,W^\dagger$ is diagonal in the basis formed by tensor products of Bell states, hence the eigenvalues can be directly determined from the expression (\ref{psumid}).
All the projectors contained in the first three terms of the right hand side of (\ref{psumid}) determine the $2^{n+1}-1$ dimensional zero eigenvalue subspace
and it is easy to see that on the remaining subspace the eigenvalue is one. Thus, all the eigenvalues of $WR\,W^\dagger$ are nonnegative, which proves that $R$ is a positive semidefinite operator.
This proves that the fidelity bound (\ref{eq:KRKineq2}) holds also for the case, when the bases $\{\ket{e_j}\}_{j=1}^d$, $\{\ket{f_k}\}_{k=1}^d$ are computational basis and its Hadamard transform on every qubit.

\end{document}